\documentclass[12pt]{iopart}

\usepackage{iopams,graphicx}
\begin{document}

\title[Overlap distribution of $(2+1)$-dimensional directed polymer in random media]{Numerical calculation of overlap distribution of $(2+1)$-dimensional directed polymer in random media}

\author{Masahiko Ueda}
\address{Department of Systems Science, Kyoto University, Kyoto, 606-8501, Japan}
\ead{ueda.masahiko.5r@kyoto-u.ac.jp}

\begin{abstract}
We investigate $(2+1)$-dimensional discretized directed polymers in Gaussian random media.
By numerically calculating the probability distribution function of overlap between two independent and identical systems on a common random potential, we show that there is no replica symmetry breaking.
We also show that while the mean-squared displacement of one polymer end exhibits superdiffusive behavior, the mean-squared relative distance of two polymer ends exhibits subdiffusive behavior.
\end{abstract}

%Uncomment for PACS numbers title message
%\pacs{05.40.-a, 05.70.Fh, 64.70.P-}
% Keywords required only for MST, PB, PMB, PM, JOA, JOB? 
%\vspace{2pc}
%\noindent{\it Keywords}: Article preparation, IOP journals
% Uncomment for Submitted to journal title message
%\submitto{\JPA}
% Comment out if separate title page not required

\maketitle

\section{Introduction}
\label{sec:intro}
Directed polymer in random media is one of the simplest models of disordered systems in statistical mechanics \cite{HalZha1995}.
Because it is related to the Kardar-Parisi-Zhang (KPZ) equation describing out-of-equilibrium surface growth \cite{KPZ1986}, it has frequently been investigated.
While $(1+1)$-dimensional directed polymer has been well-studied both theoretically and experimentally \cite{TSSS2011}, properties of directed polymers in higher dimension are not yet well-understood.
One of the unsolved problems is whether there is replica symmetry breaking (RSB) in directed polymers in higher dimension.
RSB is a concept describing the low-temperature phase of mean-field spin glass models in equilibrium state, and implies that a system has several stable disordered states \cite{MPV1987}.
The existence of RSB is detected by non-trivial distribution of overlap between two independent and identical systems.
It has been known that $(d+1)$-dimensional directed polymers with $d\leq 2$ are in strong disorder phase in any finite temperature \cite{Mez1990,CarHu2002,CSY2003}.
In contrast, in dimension $d>2$, there is a phase transition between the low-temperature localized phase and the high-temperature free phase \cite{ImbSpe1998,KBM1991,Muk1994}.
Although we know the system is localized in the low-temperature phase in both cases, whether RSB occurs or not (that is, whether the number of stable states in the localized phase is several or one) is not known.
It should be noted that this cannot be distinguished by only average overlap, and the whole information of overlap distribution is needed.
Because theoretical calculation for higher dimensional cases is difficult, we need to rely on numerical calculation.

In this study, we investigate the overlap distribution of $(2+1)$-dimensional directed polymers in Gaussian random media \cite{Hal2012,OAF2013,Hal2013,OKG2014,HalPal2014,PagPar2015,AFRO2015,KOG2016,SinNan2016}.
In our previous study \cite{UedSas2015}, we proposed numerical method to calculate overlap distribution of directed polymers, and calculated overlap distribution of $(1+1)$-dimensional directed polymer in random media, where we found that while discretized model does not exhibit RSB, continuous model seems to exhibit RSB.
Here we apply this method to $(2+1)$-dimensional discretized model, and numerically show that there is no RSB (the number of stable states is one) in this model.

The paper is organized as follows.
In section \ref{sec:model}, we introduce the model.
In section \ref{sec:method}, we introduce overlap between trajectories, and explain how to calculate numerically overlap distribution.
In section \ref{sec:results}, we provide numerical results supporting that RSB does not occur.
In this section, we also calculate the mean-squared displacement and the mean-squared relative distance, which exhibit different diffusive behavior.
Section \ref{sec:conclusion} is devoted to concluding remarks.

\section{Model}
\label{sec:model}
We consider a polymer on a $(2+1)$-dimensional lattice as a path $[\textrm{\boldmath $x$}]\equiv(\textrm{\boldmath $x$}_0,\textrm{\boldmath $x$}_1,\cdots,\textrm{\boldmath $x$}_t)$ of length $t$ with $\textrm{\boldmath $x$}_j =\left( x_j, y_j \right) \in \{1, \cdots, L\}\times \{1, \cdots, L\}$ satisfying the constraint 
\begin{eqnarray}
 \textrm{\boldmath $x$}_{j+1}\in\left\{ \textrm{\boldmath $x$}_j-\textrm{\boldmath $e$}_x,\textrm{\boldmath $x$}_j-\textrm{\boldmath $e$}_y, \textrm{\boldmath $x$}_j,\textrm{\boldmath $x$}_j+\textrm{\boldmath $e$}_x, \textrm{\boldmath $x$}_j+\textrm{\boldmath $e$}_y \right\},
\end{eqnarray}
where $\textrm{\boldmath $e$}_x$ and $\textrm{\boldmath $e$}_y$ are unit vectors in $x$ and $y$ directions, respectively.
The periodic boundary conditions are imposed in the $x$ and $y$ directions.
The polymer is subjected to a zero mean Gaussian random potential $v(\textrm{\boldmath $x$},j)$ with variance 
\begin{eqnarray}
 \overline{v(\textrm{\boldmath $x$},j) v(\textrm{\boldmath $x$}^\prime,j^\prime)} = A\delta_{x,x^\prime}\delta_{y,y^\prime}\delta_{j,j^\prime}.
\end{eqnarray}
We assume the following equilibrium distribution of the polymer with one end fixed:
\begin{eqnarray}
\fl \mathcal{P}_\mathrm{DP}\left[\textrm{\boldmath $x$}|\textrm{\boldmath $x$}_0=\textrm{\boldmath $a$}\right] 
&=& \frac{1}{Z} \prod_{j=0}^{t-1} e^{- v\left( \textrm{\boldmath $x$}_{j+1},j+1 \right)} 
\left[ \delta_{\textrm{\boldmath $x$}_{j+1},\textrm{\boldmath $x$}_{j}} + \gamma \delta_{\textrm{\boldmath $x$}_{j+1},\textrm{\boldmath $x$}_{j}+\textrm{\boldmath $e$}_x} \right. \nonumber \\
\fl && \qquad \left. + \gamma \delta_{\textrm{\boldmath $x$}_{j+1},\textrm{\boldmath $x$}_{j}-\textrm{\boldmath $e$}_x} + \gamma \delta_{\textrm{\boldmath $x$}_{j+1},\textrm{\boldmath $x$}_{j}+\textrm{\boldmath $e$}_y} + \gamma \delta_{\textrm{\boldmath $x$}_{j+1},\textrm{\boldmath $x$}_{j}-\textrm{\boldmath $e$}_y} \right],
\label{eq:pp_ddp}
\end{eqnarray}
where $Z$ is the normalization constant, $\textrm{\boldmath $a$}$ is some site on a lattice, and $\gamma$ is a parameter related to an elastic constant \cite{Mez1990}.
The partition function $Z$ of this model was calculated by using a method of the transfer matrix \cite{HalZha1995}
\begin{eqnarray}
Z\left( \textrm{\boldmath $x$}, j+1 \right) &=& \sum_{\textrm{\boldmath $x$}^\prime} 
T_{j+1}\left(\textrm{\boldmath $x$}|\textrm{\boldmath $x$}^\prime \right) Z\left( \textrm{\boldmath $x$}^\prime, j \right), \label{eq:TM}\\
 T_{j+1}\left(\textrm{\boldmath $x$}|\textrm{\boldmath $x$}^\prime \right) &\equiv& 
\frac{1}{1+4\gamma}e^{- v\left( \textrm{\boldmath $x$},j+1 \right)} 
\left[ \delta_{\textrm{\boldmath $x$},\textrm{\boldmath $x$}^\prime} + \gamma \delta_{\textrm{\boldmath $x$},\textrm{\boldmath $x$}^\prime+\textrm{\boldmath $e$}_x} 
\right. \nonumber \\
&& \qquad \left. + \gamma \delta_{\textrm{\boldmath $x$},\textrm{\boldmath $x$}^\prime-\textrm{\boldmath $e$}_x} + \gamma \delta_{\textrm{\boldmath $x$},\textrm{\boldmath $x$}^\prime+\textrm{\boldmath $e$}_y} + \gamma \delta_{\textrm{\boldmath $x$},\textrm{\boldmath $x$}^\prime-\textrm{\boldmath $e$}_y} \right]
\end{eqnarray}
with the initial condition $Z\left( \textrm{\boldmath $x$}, 0 \right)=\delta_{\textrm{\boldmath $x$},\textrm{\boldmath $a$}}$.
The partition function $Z$ is obtained as $Z=\sum_\textrm{\boldmath $x$}Z\left( \textrm{\boldmath $x$}, t \right)$.
Below we describe thermal average (with respect to polymer configurations) by $\left\langle \cdots \right\rangle$, and disorder average (with respect to random potentials) by $\overline{\cdots}$.

\section{Method}
\label{sec:method}
We define the overlap between two trajectories $\left[\textrm{\boldmath $x$}^{(1)}\right]$ and $\left[\textrm{\boldmath $x$}^{(2)}\right]$ on a common random potential as
\begin{eqnarray}
q\left( \left[\textrm{\boldmath $x$}^{(1)}\right], \left[\textrm{\boldmath $x$}^{(2)}\right] \right) 
&\equiv& \frac{1}{t} \sum_{j=1}^t \delta_{\textrm{\boldmath $x$}^{(1)}_j, \textrm{\boldmath $x$}^{(2)}_j}.
\end{eqnarray}
This overlap takes $1$ if configurations of two polymers are exactly the same, and $0$ if their configurations are independent.
Therefore, overlap is an order parameter detecting localization.
We define the probability distribution function of overlap by
\begin{eqnarray}
 P(q) &=& \overline{\left\langle \delta\left( q- q\left( \left[\textrm{\boldmath $x$}^{(1)}\right], \left[\textrm{\boldmath $x$}^{(2)}\right] \right) \right) \right\rangle}.
\end{eqnarray}
When $\lim_{t\rightarrow \infty}P(q) = \delta(q)$, configurations of two polymers are not correlated.
When $\lim_{t\rightarrow \infty}P(q) = \delta(q-q_*)$ with $q_*\neq 0$, polymers are localized to one specific configuration.
These two cases are said to be replica symmetric, because two independent and identical replicas behave similarly.
When $P(q)$ has a non-trivial peak in addition to the trivial peak $\delta(q)$ in the limit $t\rightarrow \infty$, this means that there are several stable configurations.
In this case, replica symmetry is said to be broken.

We explain our calculation method of overlap distribution $P(q)$, which relies on a Markov chain model whose path probability is equivalent to (\ref{eq:pp_ddp}). 
Below, we construct the Markov chain explicitly by using retrospective process \cite{JacSol2010,UedSas2015}.
First, we define a function $\Phi(\textrm{\boldmath $x$},j)$ by an equation
\begin{eqnarray}
\Phi\left( \textrm{\boldmath $x$}^\prime,j \right) 
&=& \sum_\textrm{\boldmath $x$} \Phi\left( \textrm{\boldmath $x$},j+1 \right) T_{j+1}\left(\textrm{\boldmath $x$}|\textrm{\boldmath $x$}^\prime \right)
 \label{eq:discrete_dynamics_field}
\end{eqnarray}
with the final condition $\Phi\left( \textrm{\boldmath $x$},t \right) = 1$.
Next, by defining 
\begin{eqnarray}
\tilde{T}_{j+1}\left(\textrm{\boldmath $x$}|\textrm{\boldmath $x$}^\prime \right) 
&\equiv& \frac{\Phi\left( \textrm{\boldmath $x$},j+1 \right)}
{\Phi\left( \textrm{\boldmath $x$}^\prime,j \right)} T_{j+1}\left(\textrm{\boldmath $x$}|\textrm{\boldmath $x$}^\prime \right),
\label{eq:T_tilde}
\end{eqnarray}
we can confirm $\sum_\textrm{\boldmath $x$}\tilde{T}_{j+1}\left(\textrm{\boldmath $x$}|\textrm{\boldmath $x$}^\prime \right)=1$ from the definition of $\Phi$. 
Thus, by identifying $\tilde T$ as a time dependent transition probability, we have a Markov chain
\begin{eqnarray}
P\left( \textrm{\boldmath $x$},j+1 \right) 
&=& \sum_{\textrm{\boldmath $x$}^\prime} \tilde{T}_{j+1}
\left(\textrm{\boldmath $x$}|\textrm{\boldmath $x$}^\prime \right) P\left( \textrm{\boldmath $x$}^\prime,j \right).
\label{eq:discrete_dynamics}
\end{eqnarray}
We fix the initial condition as $P(\textrm{\boldmath $x$},0)=\delta_{\textrm{\boldmath $x$},\textrm{\boldmath $a$}}$ with some $\textrm{\boldmath $a$}$.
Now, we write a realization of state at time $j$ as $\textrm{\boldmath $x$}_j$ and denote a path by $[\textrm{\boldmath $x$}]\equiv(\textrm{\boldmath $x$}_0,\textrm{\boldmath $x$}_1,\cdots,\textrm{\boldmath $x$}_t)$.
The probability of path is given by
\begin{eqnarray}
\mathcal{P}\left[\textrm{\boldmath $x$}|\textrm{\boldmath $x$}_0=\textrm{\boldmath $a$}\right] 
&\equiv& \prod_{j=0}^{t-1} \tilde{T}_{j+1}\left(\textrm{\boldmath $x$}_{j+1}|\textrm{\boldmath $x$}_j \right).
 \label{eq:bpp_dkpz}
\end{eqnarray}
This can be rewritten as
\begin{eqnarray}
\fl \mathcal{P}\left[\textrm{\boldmath $x$}|\textrm{\boldmath $x$}_0=\textrm{\boldmath $a$}\right] 
&=& \frac{\Phi\left( \textrm{\boldmath $x$}_t,t \right)}
{\Phi\left( \textrm{\boldmath $x$}_0,0 \right)}\prod_{j=0}^{t-1} 
T_{j+1}\left(\textrm{\boldmath $x$}_{j+1}|\textrm{\boldmath $x$}_j \right) \nonumber \\
\fl &=& \frac{1}{\Phi\left( \textrm{\boldmath $a$},0 \right)}
\prod_{j=0}^{t-1} T_{j+1}\left(\textrm{\boldmath $x$}_{j+1}|\textrm{\boldmath $x$}_j \right) \nonumber \\
\fl &=& \frac{1}{Z} \prod_{j=0}^{t-1} e^{- v\left(\textrm{\boldmath $x$}_{j+1},j+1 \right)} 
\left[ \delta_{\textrm{\boldmath $x$}_{j+1},\textrm{\boldmath $x$}_{j}} + \gamma \delta_{\textrm{\boldmath $x$}_{j+1},\textrm{\boldmath $x$}_{j}+\textrm{\boldmath $e$}_x} \right. \nonumber \\
\fl && \left. \qquad + \gamma \delta_{\textrm{\boldmath $x$}_{j+1},\textrm{\boldmath $x$}_{j}-\textrm{\boldmath $e$}_x} + \gamma \delta_{\textrm{\boldmath $x$}_{j+1},\textrm{\boldmath $x$}_{j}+\textrm{\boldmath $e$}_y} 
+ \gamma \delta_{\textrm{\boldmath $x$}_{j+1},\textrm{\boldmath $x$}_{j}-\textrm{\boldmath $e$}_y} \right],
\end{eqnarray}
where $Z$ is the normalization constant.
This path probability is equivalent to (\ref{eq:pp_ddp}).
In numerical calculation, $\Phi\left( \textrm{\boldmath $x$},j \right)$ is calculated by (\ref{eq:discrete_dynamics_field}), and then $\mathcal{P}\left[\textrm{\boldmath $x$}|\textrm{\boldmath $x$}_0=\textrm{\boldmath $a$}\right]$ is obtained by (\ref{eq:bpp_dkpz}), by generating trajectories according to the transition probability (\ref{eq:T_tilde}).
By collecting trajectories numerically, we obtained $P(q)$.

In numerical simulation, we set parameters $L=1024$, $A=1.0$, and $\gamma=0.1$.
We use $80000$ trajectories to calculate thermal average, and $10000$ samples to calculate disorder average.
We use eight values of $t$: $t=64, 128, 192, 256, 320, 384, 448, 512$.

\section{Numerical results}
\label{sec:results}
\begin{figure}[t]
 \includegraphics[clip, width=8.0cm]{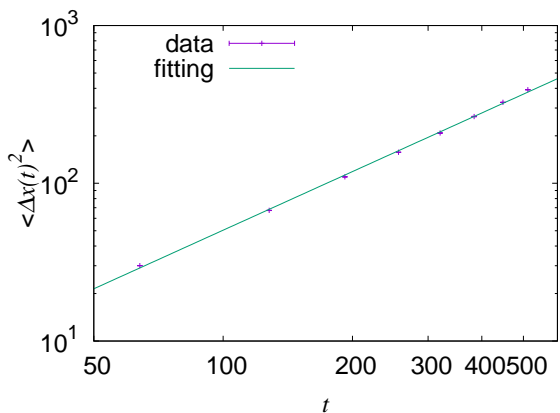}
 \caption{$t$ dependence of the mean-squared displacement $\overline{\left\langle \Delta \textrm{\boldmath $x$}(t)^2 \right\rangle}$ in a log--log plot. The errorbars are standard error. The slope of the straight line is about $1.23$.}
 \label{fig:MSD}
\end{figure}

Before calculating overlap distribution $P(q)$, we numerically calculate standard quantities characterizing diffusive behavior of one polymer end $\textrm{\boldmath $x$}_t$.
First, we study the mean-squared displacement $\overline{\left\langle \Delta \textrm{\boldmath $x$}(t)^2 \right\rangle}$ with $\Delta \textrm{\boldmath $x$}(t) \equiv \textrm{\boldmath $x$}_t-\textrm{\boldmath $x$}_0$.
In Fig. \ref{fig:MSD}, we display the mean-squared displacement for various $t$.
We can find superdiffusive behavior $\overline{\left\langle \Delta \textrm{\boldmath $x$}(t)^2 \right\rangle} \sim t^\alpha$ $(\alpha>1)$.
The value of $\alpha$ is estimated by fitting a linear equation to the log-log plot by the least squares method.
We obtain $\alpha = 1.23 \pm 0.02$, which is consistent with previous studies, because the dynamical exponent $z$ is calculated as $z=2/\alpha$.
(See Ref. \cite{PagPar2015} and its references.)
Since $\overline{\left\langle \Delta \textrm{\boldmath $x$}(t)^2 \right\rangle}$ is much smaller than $L^2$, we expect that finite size effects do not affect the result.

\begin{figure}[t]
 \includegraphics[clip, width=8.0cm]{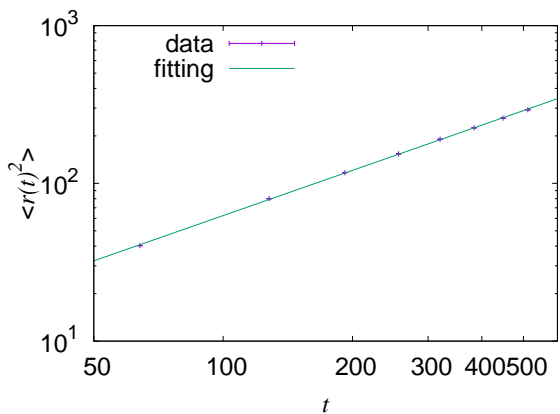}
 \caption{$t$ dependence of the mean-squared relative distance $\overline{\left\langle \textrm{\boldmath $r$}(t)^2 \right\rangle}$ in a log--log plot. The errorbars are standard error. The slope of the straight line is about $0.952$.}
 \label{fig:MSRD}
\end{figure}

Next, we study the mean-squared relative distance \cite{KlyTat1974,UedSas2015,Ued2016,SinBar2018-1,UedSas2018,SinBar2018-2}.
We consider two polymers $\left[\textrm{\boldmath $x$}^{(1)}\right]$ and $\left[\textrm{\boldmath $x$}^{(2)}\right]$ on a common random potential.
We define the mean-squared relative distance by $\overline{\left\langle \textrm{\boldmath $r$}(t)^2 \right\rangle}$ with $\textrm{\boldmath $r$}(t) \equiv \textrm{\boldmath $x$}^{(1)}_t-\textrm{\boldmath $x$}^{(2)}_t$.
In Fig. \ref{fig:MSRD}, we display the mean-squared relative distance for various $t$.
We can see slightly subdiffusive behavior $\overline{\left\langle \textrm{\boldmath $r$}(t)^2 \right\rangle} \simeq t^{\alpha_\mathrm{r}}$ $(\alpha_\mathrm{r} < 1)$, which is different from superdiffusive behavior of the mean-squared displacement $\overline{\left\langle \Delta \textrm{\boldmath $x$}(t)^2 \right\rangle}$.
We also estimate $\alpha_\mathrm{r}$ as $\alpha_\mathrm{r} = 0.952 \pm 0.005$ by fitting.
This suppression of relative diffusion is one of the evidence of localization in relative distance.
It should be noted that this subdiffusive behavior of relative distance is in contrast to $(1+1)$-dimensional (continuous) case \cite{UedSas2015}, where relative diffusion is normal $(\alpha_\mathrm{r}=1)$.

\begin{figure}[t]
 \includegraphics[clip, width=8.0cm]{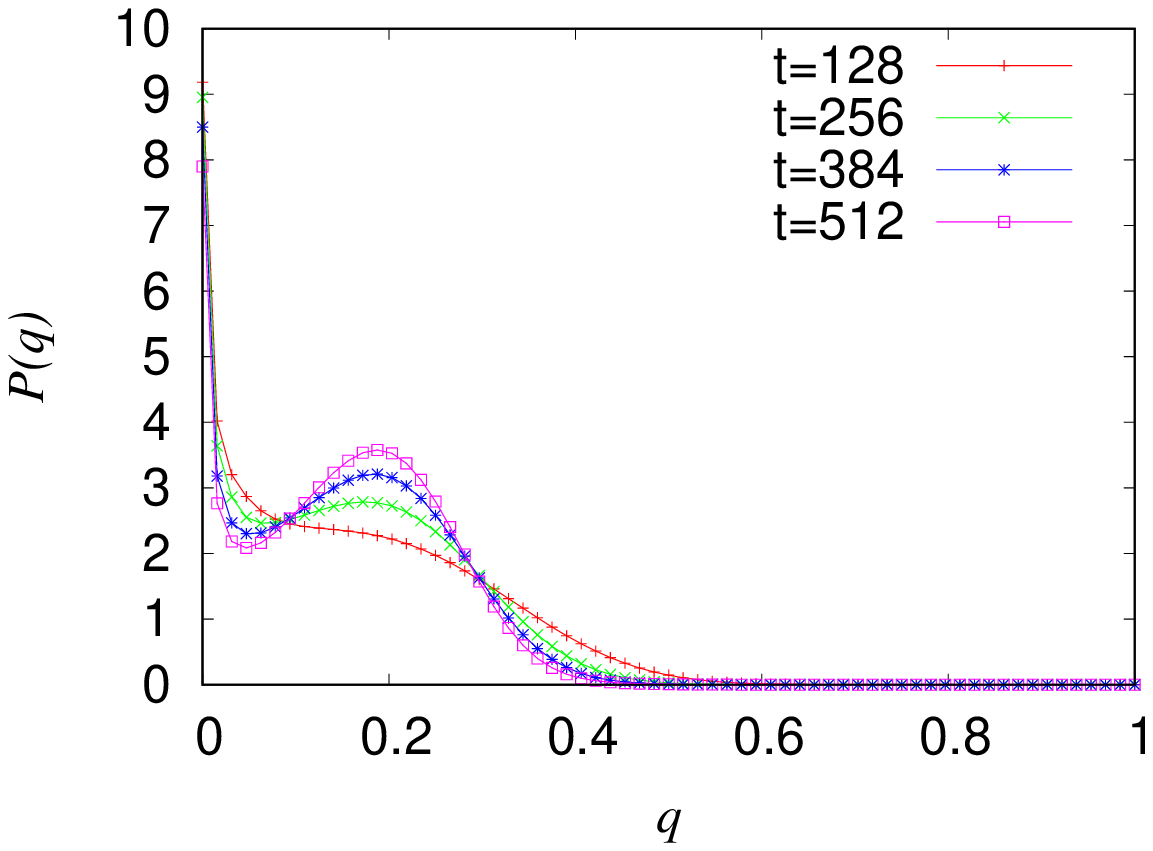}
 \includegraphics[clip, width=8.0cm]{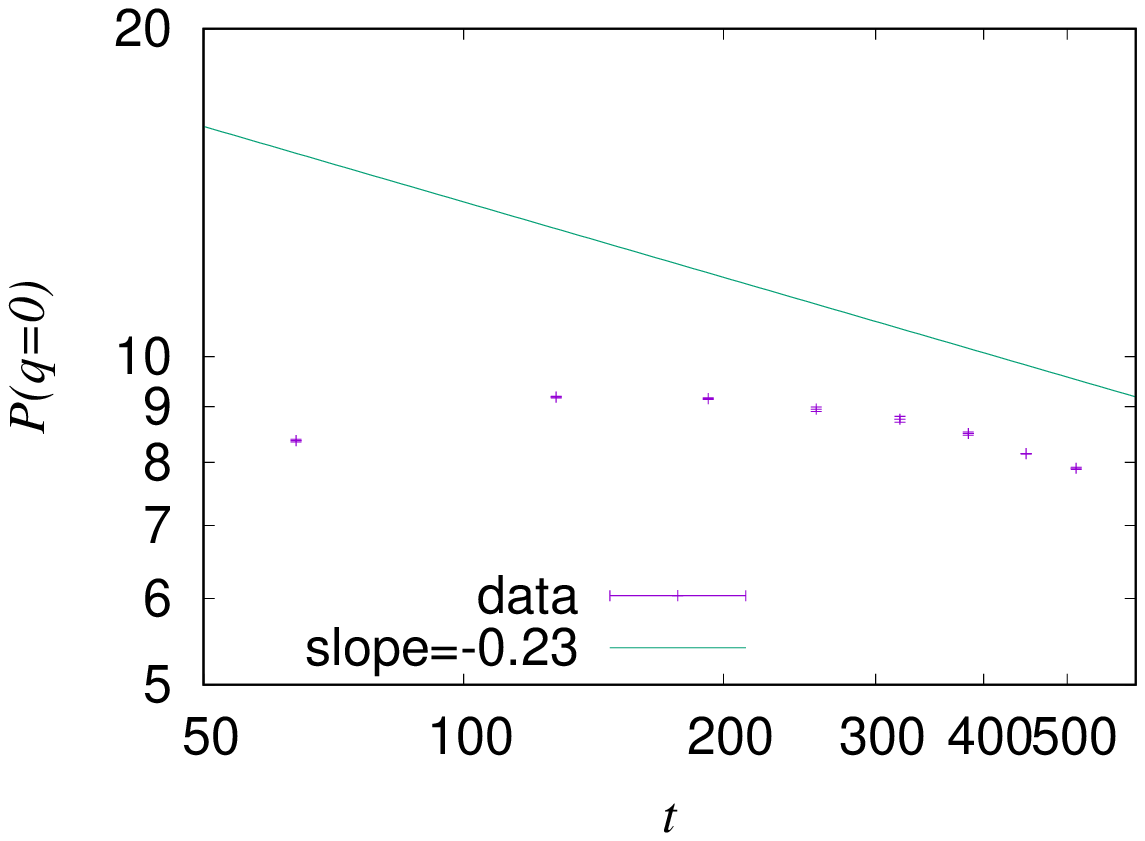}
 \caption{(Left) The probability distribution function $P(q)$ of overlap for various $t$. (Right) $t$ dependence of $P(q=0)$ in a log--log plot. The errorbars are standard error. The slope of the straight line is $-0.23$.}
 \label{fig:ovhist}
\end{figure}

Finally, we numerically calculate overlap distribution $P(q)$ \cite{UedSas2015,UedSas2017,Ued2018}.
In the left side of Fig. \ref{fig:ovhist}, we display the overlap distribution for various $t$.
We can see that the peak at $q=0$ decays with $t$ and the non-trivial peak becomes sharper and sharper with $t$.
This result suggests that there is no RSB.

In order to confirm this fact, we also focus on $t$ dependence of $P(q=0)$, which is plotted in the right side of Fig. \ref{fig:ovhist}.
For the $(1+1)$-dimensional model, it is conjectured that $P(q=0) \sim t^{-1/3}$ \cite{Mez1990,Par1990}.
This conjecture is based on the fact that $\overline{\left\langle \left[ x(t) - \left\langle x(t) \right\rangle \right]^2 \right\rangle} \sim t$ with $\overline{\left\langle \left[ x(t) - x(0) \right]^2 \right\rangle} \sim t^{4/3}$ and therefore the probability that two independent and identical systems belong to different pure states from each other decays as $t^{-1/3}$.
Because we can also theoretically prove $\overline{\left\langle \left[ \textrm{\boldmath $x$}(t) - \left\langle \textrm{\boldmath $x$}(t) \right\rangle \right]^2 \right\rangle} \sim t$ for higher dimensional cases by using the same argument as in Ref. \cite{Mez1990}, we conjecture
\begin{eqnarray}
 P(q=0) &\sim& \frac{1}{t^{\alpha-1}},
\end{eqnarray}
where $\alpha$ is calculated from the mean-squared displacement.
As we can see in the right side of Fig. \ref{fig:ovhist}, this conjecture is consistent with our numerical result when we focus on $t\geq 384$.
Therefore, we conclude that there is no RSB, and the finite temperature phase of $(2+1)$-dimensional discretized directed polymers is dominated by only one state.
Because it is known that $(2+1)$-dimensional directed polymer is localized for any finite temperature \cite{HalZha1995,CarHu2002,CSY2003}, this result is expected to hold for other parameter regions.

\section{Concluding remarks}
\label{sec:conclusion}
In this paper, we numerically calculated overlap distribution of $(2+1)$-dimensional discretized directed polymers in zero-mean Gaussian random potentials.
We found that $P(q)$ seems to converge to $\delta(q-q_*)$ with $q_*\neq 0$ in the limit $t\rightarrow \infty$, which means that there is no RSB.
We also found that while the mean-squared displacement of one polymer end exhibits superdiffusive behavior, the mean-squared relative distance of two polymer ends exhibits slightly subdiffusive behavior, which is also one evidence of localization.

Before ending this paper, we make four remarks.
The first remark is related to our numerical method.
The most-used numerical method for directed polymers in random media is Monte Carlo method \cite{Yos1996}.
Generally, in the Monte Carlo method, ruggedness of an energy landscape leads to slow relaxation to the equilibrium distribution.
In contrast, in our numerical method, because we directly generate the equilibrium ensemble of trajectories according to (\ref{eq:bpp_dkpz}), the slow-relaxation problem is avoided.
The drawback of our method is that we cannot perform numerical simulation for large $t$.
Since we need to calculate exponentially growing function of $t$ by the recurrence relation (\ref{eq:discrete_dynamics_field}), which is the product of transfer matrices, we cannot perform calculation for large $t$.
Therefore, we will need to use standard Monte Carlo method for investigating large-$t$ behavior.

The second remark is relation to population genetics.
It has been known that the transfer matrix calculation (\ref{eq:TM}) can be interpreted as population dynamics equation in population genetics \cite{Leu1987,GBM2001,KLG2006}, where $\textrm{\boldmath $x$}$ is genotype, $j$ is generation, $Z\left( \textrm{\boldmath $x$},t \right)$ is the number of individuals with genotype $\textrm{\boldmath $x$}$ at generation $t$, and $ e^{- v\left( \textrm{\boldmath $x$},j \right)}$ is growth rate.
In terms of population genetics, our Markov chain in section \ref{sec:method} is equivalent to retrospective process in phylogenetic tree \cite{SKTA2015}.
Therefore, our result implies that when genotype space is two-dimensional and fitness obeys Gaussian distribution, condensation to a specific path in phylogenetic tree occurs.

The third remark is related to the subdiffusive behavior of relative distance.
As discussed in section \ref{sec:results}, by using the same argument as in Ref. \cite{Mez1990}, we can obtain $\overline{\left\langle \left[ \textrm{\boldmath $x$}(t) - \left\langle \textrm{\boldmath $x$}(t) \right\rangle \right]^2 \right\rangle} \sim t$ for (continuous) directed polymers.
If we expect that $\overline{\left\langle \left[ \textrm{\boldmath $x$}(t) - \left\langle \textrm{\boldmath $x$}(t) \right\rangle \right]^2 \right\rangle}$ and $\overline{\left\langle \textrm{\boldmath $r$}(t)^2 \right\rangle}$ behave in the same manner, $\alpha_\mathrm{r}=1$ is expected.
Therefore, our result about subdiffusive behavior of relative distance may come from the fact that $t$ is not sufficiently large.
This point should be studied more extensively in future.

The final remark is related to difference between discretized model and continuous model.
In $(1+1)$-dimensional case, properties of discretized model and continuous model seem to be different \cite{UedSas2015}: while the former exhibits decay of $P(q=0)$, the latter exhibits increase of $P(q=0)$.
Investigating the existence of RSB in $(2+1)$-dimensional continuous directed polymer is a future problem.

% acknowledgements
\ack
This study was supported by JSPS KAKENHI Grant Numbers JP18H06476.

%\bibliographystyle{iopart-num}
%\bibliography{DP2d_bib} 

\section*{References}
\begin{thebibliography}{99}

 % directed polymer
 \bibitem{HalZha1995} Halpin-Healy T and Zhang YC, 1995 {\it Phys. Rep.} {\bf 254} 215
 
 % KPZ
 \bibitem{KPZ1986} Kardar M, Parisi G and Zhang YC, 1986 {\it Phys. Rev. Lett.} {\bf 56} 889
 
 % KPZ 1-dimension
 \bibitem{TSSS2011} Takeuchi KA, Sano M, Sasamoto T and Spohn H, 2011 {\it Sci. Rep.} {\bf 1} 34
 
 % RSB
 \bibitem{MPV1987} M\'{e}zard M, Parisi G and Virasoro M A, 1987 {\it Spin glass theory and beyond} (Singapore: World Scientific)
 
 % DP low-dimensional
 \bibitem{Mez1990} M\'{e}zard M, 1990 {\it J. Phys. France} {\bf 51} 1831
 \bibitem{CarHu2002} Carmona P and Hu Y, 2002 {\it Probab. Theory Relat. Fields} {\bf 124} 431
 \bibitem{CSY2003} Comets F, Shiga T and Yoshida N, 2003 {\it Bernoulli} {\bf 9} 705
 
 % DP high-dimensional
 \bibitem{ImbSpe1998} Imbrie JZ and Spencer T, 1988 {\it J. Stat. Phys.} {\bf 52} 609
 \bibitem{KBM1991} Kim JM, Bray AJ and Moore MA, 1991 {\it Phys. Rev. A} {\bf 44} R4782
 \bibitem{Muk1994} Mukherji S, 1994 {\it Phys. Rev. E} {\bf 50} R2407
 
 % DP (2+1)
 \bibitem{Hal2012} Halpin-Healy T, 2012 {\it Phys. Rev. Lett.} {\bf 109} 170602
 \bibitem{OAF2013} Oliveira TJ, Alves SG and Ferreira SC, 2013 {\it Phys. Rev. E} {\bf 87} 040102(R)
 \bibitem{Hal2013} Halpin-Healy T, 2013 {\it Phys. Rev. E} {\bf 88} 042118
 \bibitem{OKG2014} \'{O}dor G, Kelling J and Gemming S, 2014 {\it Phys. Rev. E} {\bf 89} 032146
 \bibitem{HalPal2014} Halpin-Healy T and Palasantzas G, 2014 {\it Europhys. Lett.} {\bf 105} 50001
 \bibitem{PagPar2015} Pagnani A and Parisi G, 2015 {\it Phys. Rev. E} {\bf 92} 010101(R)
 \bibitem{AFRO2015} Almeida RAL, Ferreira SO, Ribeiro IRB and Oliveira TJ, 2015 {\it Europhys. Lett.} {\bf 109} 46003
 \bibitem{KOG2016} Kelling J, \'{O}dor G and Gemming S, 2016 {\it Phys. Rev. E} {\bf 94} 022107
 \bibitem{SinNan2016} Singha T and Nandy MK, 2016 {\it J. Stat. Mech.: Theor. Exp.} {\bf 2016} 103204
 
 % RSB in trajectory
 \bibitem{UedSas2015} Ueda M and Sasa S, 2015 {\it Phys. Rev. Lett.} {\bf 115} 080605
 
 % retrospective process
 \bibitem{JacSol2010} Jack RL and Sollich P, 2010 {\it Prog. Theor. Phys. Suppl.} {\bf 184} 304
 
 % relative diffusion
 \bibitem{KlyTat1974} Klyatskin VI and Tatarskii VI, 1974 {\it Sov. Phys. Usp.} {\bf 16} 494
 \bibitem{Ued2016} Ueda M, 2016 {\it J. Stat. Mech.: Theor. Exp.} {\bf 2016} 023206
 \bibitem{SinBar2018-1} Singha T and Barma M, 2018 {\it Phys. Rev. Lett.} {\bf 121} 128901
 \bibitem{UedSas2018} Ueda M and Sasa S, 2018 {\it Phys. Rev. Lett.} {\bf 121} 128902
 \bibitem{SinBar2018-2} Singha T and Barma M, 2018 {\it Phys. Rev. E} {\bf 98} 052148
 
 % overlap between trajectories
 \bibitem{UedSas2017} Ueda M and Sasa S, 2017 {\it J. Phys. A: Math. Theor.} {\bf 50} 125001
 \bibitem{Ued2018} Ueda M, 2018 {\it J. Stat. Mech.: Theor. Exp.} {\bf 2018} 053304
 
 % DP (1+1)
 \bibitem{Par1990} Parisi G, 1990 {\it J. Phys. France} {\bf 51} 1595
 
 % DP Monte Carlo
 \bibitem{Yos1996} Yoshino H, 1996 {\it J. Phys. A: Math. Gen.} {\bf 29} 1421
 
 % population genetics
 \bibitem{Leu1987} Leuth\"{a}usser I, 1987 {\it J. Stat. Phys.} {\bf 48} 343
 \bibitem{GBM2001} Giardina I, Bouchaud JP and M\'{e}zard M, 2001 {\it J. Phys. A: Math. Gen.} {\bf 34} L245
 \bibitem{KLG2006} Kussell E, Leibler S and Grosberg A, 2006 {\it Phys. Rev. Lett.} {\bf 97} 068101
 
 % retrospective process in population genetics
 \bibitem{SKTA2015} Sughiyama Y, Kobayashi TJ, Tsumura K and Aihara K, 2015 {\it Phys. Rev. E} {\bf 91} 032120
 
\end{thebibliography}

\end{document}